\documentstyle[epsf]{elsart}

\begin{document}

\begin{frontmatter}

\title{Evidence for quantum chaos in the plasma phase of QCD}

\author[Wien]{R.~Pullirsch}, 
\author[Wien]{K.~Rabitsch},
\author[Munich]{T.~Wettig}, and
\author[Wien]{H.~Markum}
 
\address[Wien]{Institut f\"ur Kernphysik, Technische Universit\"at
          Wien, A-1040 Wien, Austria}
\address[Munich]{Institut f\"ur Theoretische Physik, Technische
          Universit\"at M\"unchen, D-85747 Garching, Germany}
       
\begin{abstract}
  We investigate the eigenvalue spectrum of the staggered Dirac matrix
  in SU(3) gauge theory and in full QCD on a $6^3\times 4$ lattice.
  As a measure of the fluctuation properties of the eigenvalues, we
  study the nearest-neighbor spacing distribution $P(s)$ for various
  values of $\beta$ both in the confinement and in the deconfinement
  phase.  In both phases except far into the deconfinement region, the
  lattice data agree with the Wigner surmise of random-matrix theory
  which is indicative of quantum chaos.  We do not find signs of a
  transition to Poisson regularity at the deconfinement phase
  transition. 
\end{abstract}

\end{frontmatter}

\maketitle

\section{Introduction}
\label{sec1}

The properties of the eigenvalues of the Dirac operator are of great
importance for the understanding of certain features of QCD.  For
example, the accumulation of small eigenvalues is, via the
Banks-Casher formula \cite{Bank80}, related to the spontaneous
breaking of chiral symmetry.  Recently, the fluctuation properties of
the eigenvalues in the bulk of the spectrum have also attracted
attention.  It was shown in Ref.~\cite{Hala95} that on the scale of
the mean level spacing they are described by random-matrix theory
(RMT). In particular, the nearest-neighbor spacing distribution
$P(s)$, i.e., the distribution of spacings $s$ between adjacent
eigenvalues on the unfolded scale (see below), agrees with the Wigner
surmise of RMT.  According to the so-called Bohigas-conjecture
\cite{Bohi84}, quantum systems whose classical counterparts are
chaotic have a nearest-neighbor spacing distribution given by RMT
whereas systems whose classical counterparts are integrable obey a
Poisson distribution, $P(s)=e^{-s}$.  Therefore, the specific form of
$P(s)$ is often taken as a criterion for the presence or absence of
``quantum chaos''.  It should be noted, however, that there is no
universally accepted proof of the Bohigas-conjecture yet.  The field
of quantum chaos is still developing, and there are many open
theoretical problems.  For reviews on this subject we refer to
Ref.~\cite{QC}.

A transition in $P(s)$ from Wigner to Poisson behavior is observed at
the localization transition in disordered mesoscopic systems
\cite{mesosc}.  In Ref.~\cite{Hala95}, the question was raised if
there is such a transition from chaotic to regular behavior in the
case of the lattice Dirac operator, and the present study serves to
investigate this question.  For a preliminary account of results for
quenched QCD, see Ref.~\cite{Mark97}.  Recently, a Wigner to Poisson
transition was also studied in the context of a spatially homogeneous
Yang-Mills-Higgs system \cite{Sala97}.  The question if chaos in an
$N$-component $\varphi^4$-theory in the presence of an external field
survives quantization was investigated in Ref.~\cite{Gatt97}.  An
analytical treatment of the transition between Wigner and Poisson
behavior can be found in Ref.~\cite{Guhr96}.

In RMT, one has to distinguish between several universality classes
which are determined by the symmetries of the system.  For the case of
the QCD Dirac operator, this classification was done in
Ref.~\cite{Verb94}.  Depending on the number of colors and the
representation of the quarks, the Dirac operator is described by one
of the three chiral ensembles of RMT.  As far as the fluctuation
properties in the bulk of the spectrum are concerned, the predictions
of the chiral ensembles are identical to those of the ordinary
ensembles \cite{Fox64}.  In Ref.~\cite{Hala95}, the Dirac matrix was
studied in SU(2) using both staggered and Wilson fermions which
correspond to the symplectic and orthogonal ensemble, respectively.
Here, we study SU(3) with staggered fermions which corresponds to the
chiral unitary ensemble.  We thus cover the last remaining symmetry
class in the confined phase.  Although the unitary ensemble is, from a
mathematical point of view, the simplest one, the RMT result for the
nearest-neighbor spacing distribution is still rather complicated.  It
can be expressed in terms of so-called prolate spheroidal functions,
see Ref.~\cite{Meht91} where $P(s)$ has also been tabulated.  A very
good approximation to $P(s)$ is provided by
\begin{equation}
  \label{eq1}
  P(s)=\frac{32}{\pi^2}s^2e^{-\frac{4}{\pi}s^2} \:.
\end{equation}
This is the Wigner surmise for the unitary ensemble.

\section{Eigenvalue analysis}
\label{sec2}

We generated gauge field configurations using the standard Wilson
plaquette action for SU(3) with and without dynamical fermions in the
Kogut-Susskind prescription. We have worked on a $6^3\times 4$ lattice
with various values of the inverse gauge coupling $\beta=6/g^2$ both
in the confinement and deconfinement phase. The boundary conditions
were periodic for the gluons and periodic in space and anti-periodic
in Euclidean time for the fermions.  We typically produced 10
independent equilibrium configurations (separated by 1000 sweeps) for
each $\beta$.  Because of the spectral ergodicity property of RMT one
can replace ensemble averages by spectral averages if one is only
interested in bulk properties, and thus one does not need a large
number of independent configurations to compute $P(s)$.

By definition, the Dirac operator
$\FMSlash{D}=\FMSlash{\partial}+ig\FMSlash{A}$
from which the spectrum is calculated does not contain the quark mass term.
The operator is anti-hermitian so that all eigenvalues are imaginary.  For
convenience, we denote them by $i\lambda_n$ and refer to the
$\lambda_n$ as the eigenvalues in the following.  Because of
$\{\FMSlash{D},\gamma_5\}=0$ the $\lambda_n$ occur in pairs of
opposite sign.  All spectra were checked against the analytical sum
rules
\begin{equation}
\sum_{n} \lambda_n = 0 \qquad {\rm and} \qquad
\sum_{\lambda_n>0} \lambda_n^2 = 3V \:,
\end{equation}
where V is the lattice volume. 

To construct the nearest-neighbor spacing distribution $P(s)$ from the
eigenvalues, one first has to ``unfold'' the spectra.  This procedure
is a local rescaling of the energy scale so that the mean level
spacing is equal to unity on the unfolded scale.  One first defines
the staircase function $N(E)$ to be the number of eigenvalues with
$\lambda\le E$. This staircase function has an average part and a
fluctuating part, $N(E)=N_{\rm av}(E)+N_{\rm fl}(E)$.  The average
part is extracted by fitting $N(E)$ to a smooth curve, e.g., to a
low-order polynomial.  One then defines the unfolded energies to be
$x_n=N_{\rm av}(E_n)$.  As a consequence, the sequence $\{x_n\}$ has
mean level spacing equal to unity.  Ensemble and spectral averages are
only meaningful after unfolding.

\section{Results and Discussion}
\label{sec3}

\begin{figure}
\begin{center}
\begin{tabular}{llcll}
  \hspace*{7mm} & \hspace*{3mm} {\large Confinement} & \hspace*{15mm} &
  \hspace*{7mm} & \hspace*{2mm} {\large Deconfinement}  \\ 
  \vspace*{-3mm}
  &$\beta=2.8$ (solid line) &&& $\beta=6.0$ (solid line) \\ 
  \vspace*{-2mm}
  &$\beta=5.0$ (dashed line) &&& $\beta=10.0$ (dashed line) \\[2mm]
  \multicolumn{2}{c}{\epsfxsize=5cm\epsffile{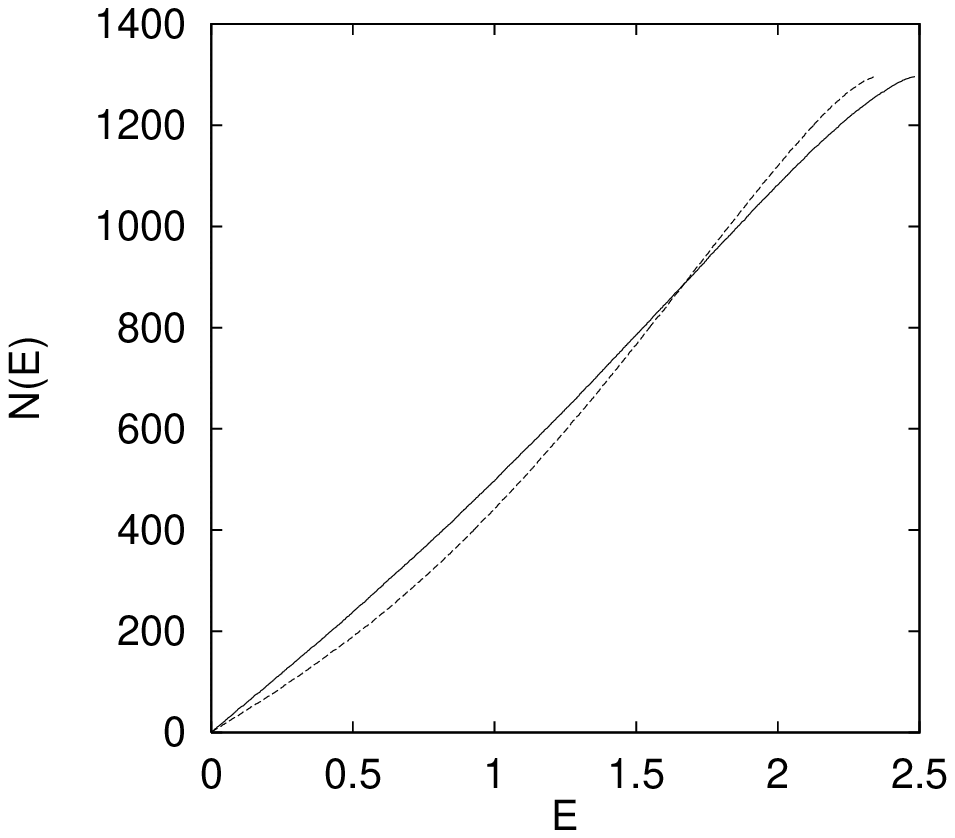}}  &&
  \multicolumn{2}{c}{\epsfxsize=5cm\epsffile{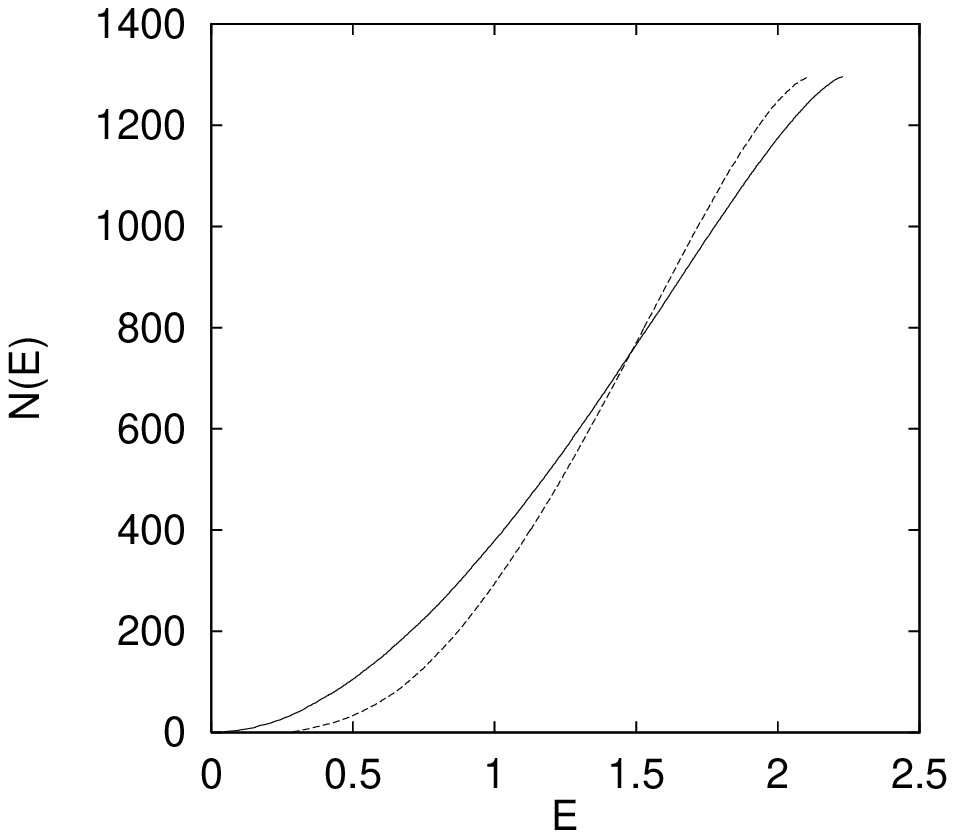}}
\end{tabular}
\end{center}
\vspace*{-1mm}
\caption{The staircase function $N(E)$ representing the number of
  positive eigenvalues $\le E$ for typical configurations on a $6^3
  \times 4$ lattice in pure SU(3).}
\label{fig1}
\end{figure}

We begin the discussion of our results with the pure gluonic case.
Figure~\ref{fig1} shows the staircase function $N(E)$ which in the
continuum is given by $N(E)=\int_0^E\rho(\lambda)d\lambda$, where
$\rho(\lambda)$ is the spectral density.  Note that $\rho(\lambda)$
cannot be obtained universally from a random-matrix model. The 
staircase function
is not very sensitive to the value of $\beta$, neither in the
confinement nor in the deconfinement regime. However, across the phase
transition $N(E)$ is diminished for small $E$ reflecting the
suppression of small eigenvalues.

\begin{figure}
\begin{center}
\begin{tabular}{ccccc}
  \hspace*{5mm} & {\large Confinement} & \hspace*{15mm}   &
  \hspace*{4mm} & {\large Deconfinement} \\
  \vspace*{-2mm}
  &$\beta=2.8$ &&& $\beta=6.0$ \\[2mm]
  \multicolumn{2}{c}{\epsfxsize=5cm\epsffile{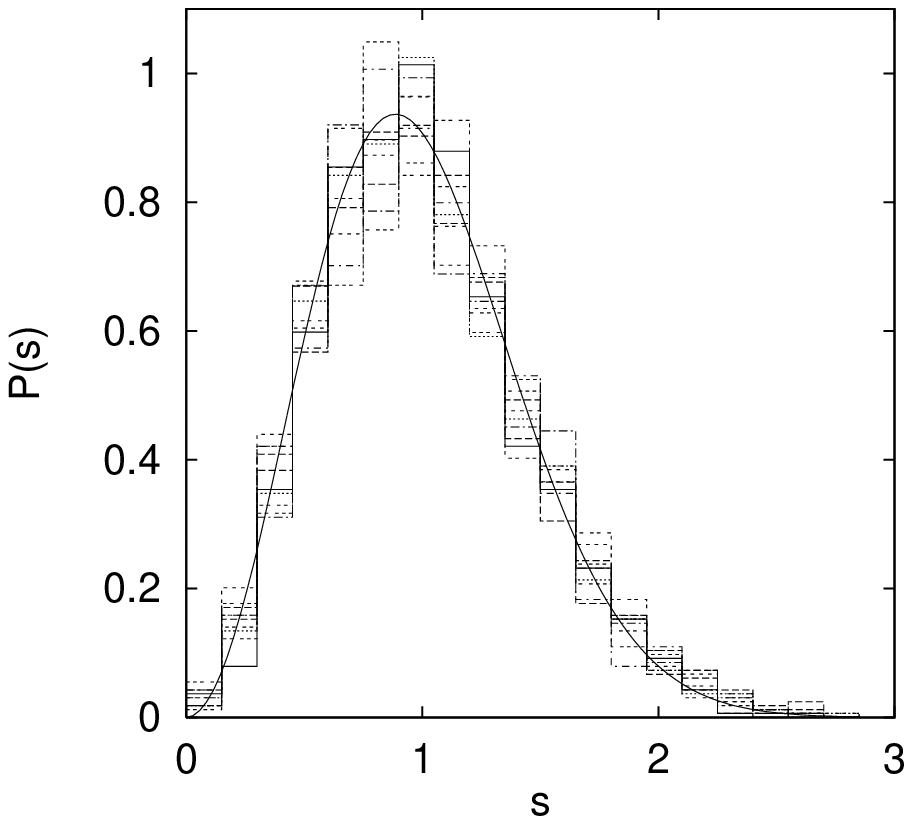}} &&
  \multicolumn{2}{c}{\epsfxsize=5cm\epsffile{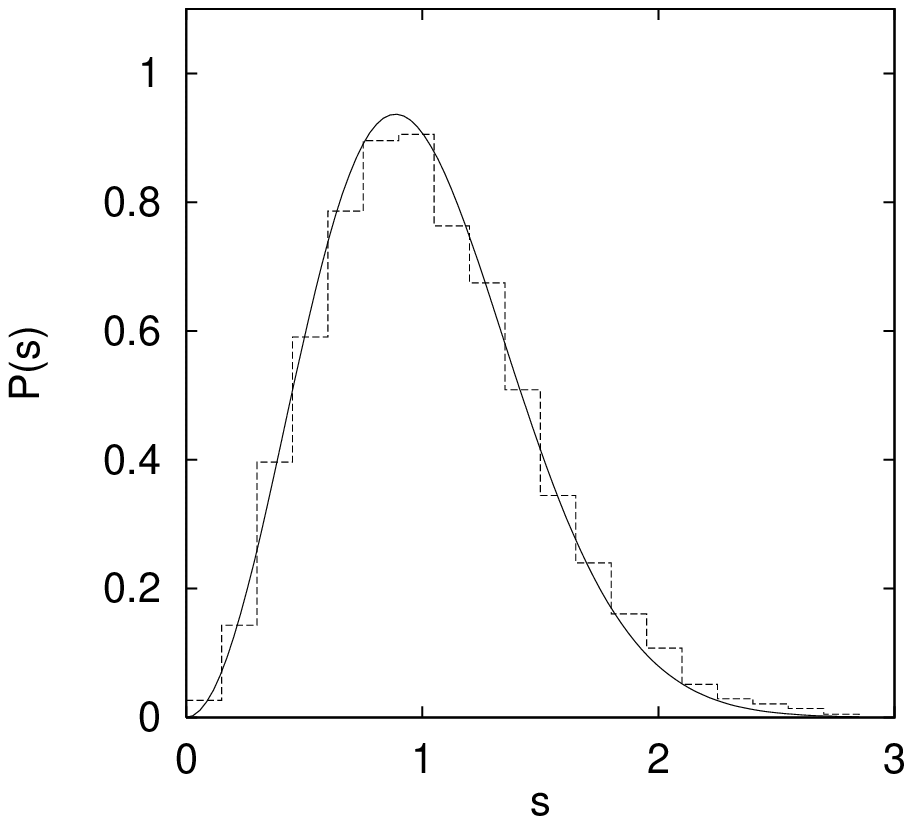}} \\[0mm]
  \vspace*{-2mm}
  & $\beta=5.0$ &&& $\beta=10.0$ \\[2mm]
  \multicolumn{2}{c}{\epsfxsize=5cm\epsffile{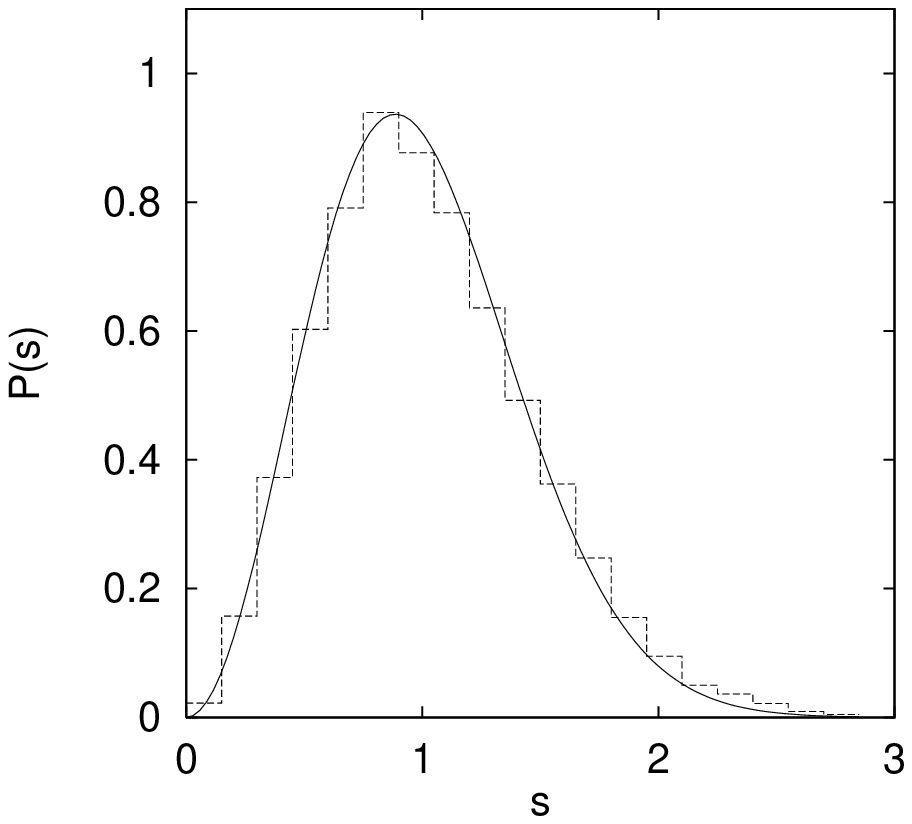}} &&
  \multicolumn{2}{c}{\epsfxsize=5cm\epsffile{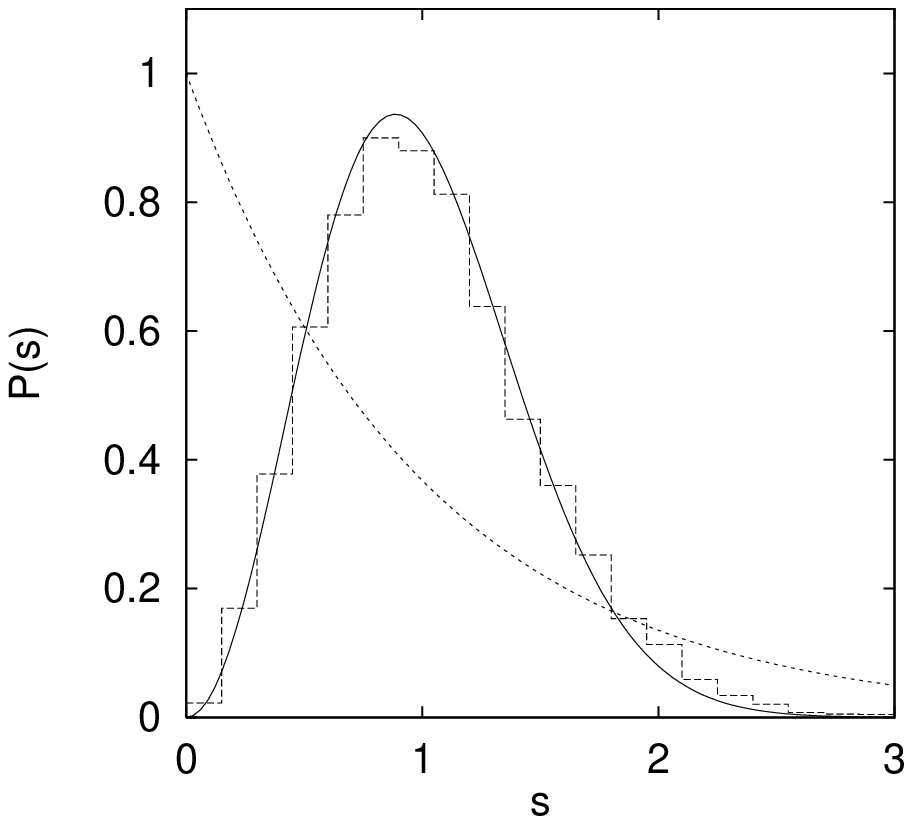}}
\end{tabular}
\end{center}
\vspace*{-1mm}
\caption{The nearest-neighbor spacing distribution $P(s)$ averaged
  over 10 independent configurations (except for $\beta=2.8$ where all
  single configurations are plotted to give an indication of the
  deviations) on a $6^3 \times 4$ lattice in pure SU(3) (histograms)
  compared with the random-matrix result (solid lines).  For
  comparison, the Poisson distribution $P(s)=e^{-s}$ is plotted for
  $\beta=10.0$ (dotted line).  There are no changes in $P(s)$ across
  the deconfinement phase transition.}
\label{fig2}
\end{figure}

Figure~\ref{fig2} shows the nearest-neighbor spacing distribution
$P(s)$ corresponding to the parameters of Fig.~\ref{fig1}, compared
with the RMT prediction.  In the confinement phase, we find the
expected agreement of $P(s)$ with the Wigner surmise of
Eq.~(\ref{eq1}).  In the deconfinement phase, we still observe
agreement with the RMT result up to $\beta=10.0$ (we have also
plotted the Poisson distribution for comparison).

\begin{figure}
\begin{center}
\begin{tabular}{llcll}
  \hspace*{10mm} & \hspace*{-5mm} {\large Confinement $\beta=5.2$} & 
  \hspace*{5mm} &
  \hspace*{10mm} & \hspace*{-8mm} {\large Deconfinement $\beta=5.4$} \\
  \vspace*{-3mm}
  & $ma=0.1$  (solid line) &&& $ma=0.1$ (solid line) \\
  \vspace*{-2mm}
  & $ma=0.05$ (dashed line) &&& $ma=0.05$ (dashed line) \\[2mm]
  \multicolumn{2}{c}{\epsfxsize=5cm\epsffile{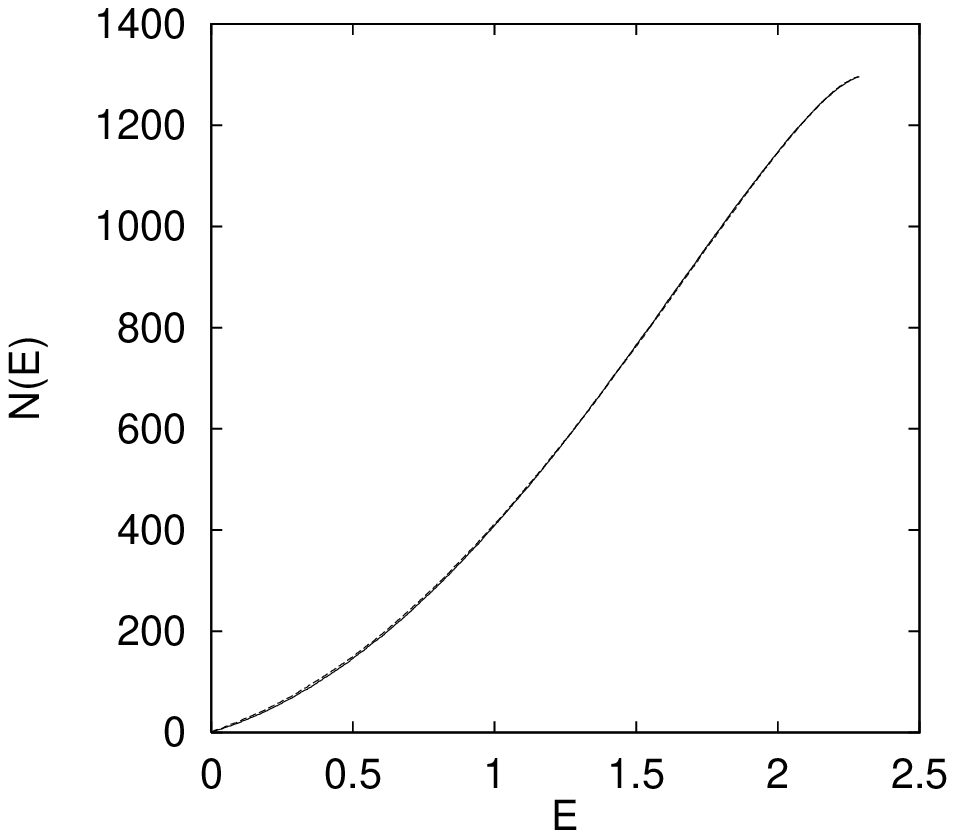}} &&
  \multicolumn{2}{c}{\epsfxsize=5cm\epsffile{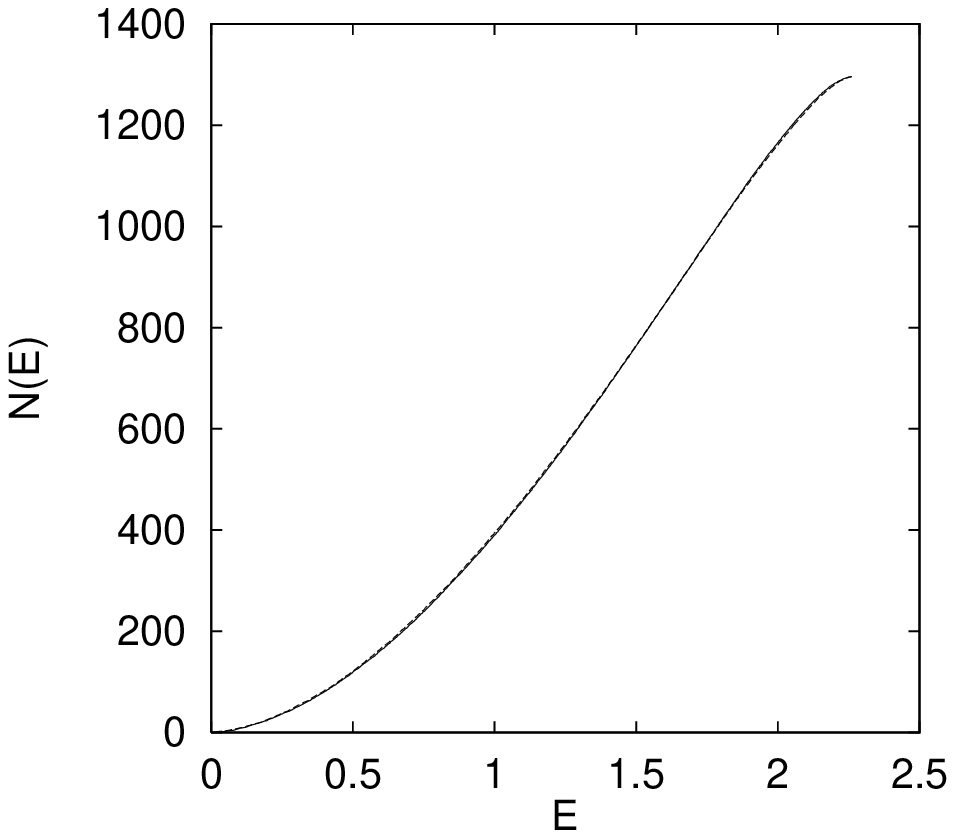}}
\end{tabular}
\end{center}
\vspace*{-1mm}
\caption{The staircase function $N(E)$ for typical configurations on
  a $6^3 \times 4$ lattice in full QCD with different quark masses.}
\label{fig3}
\end{figure}

We continue with the case of full QCD with $N_f=3$ degenerate flavors
of staggered quarks with mass $ma = 0.1$ and $m a = 0.05$,
respectively. The staircase function $N(E)$ in Fig.~\ref{fig3} shows
practically no dependence on the masses considered.

\begin{figure}
\begin{center}
\begin{tabular}{ccccc}
  & {\large Confinement $\beta=5.2$}  & \hspace*{10mm}   & &
  {\large Deconfinement $\beta=5.4$} \\
  \vspace*{-2mm}
  & $ma=0.1$ &&& $ma=0.1$ \\[2mm]
  \multicolumn{2}{c}{\epsfxsize=5cm\epsffile{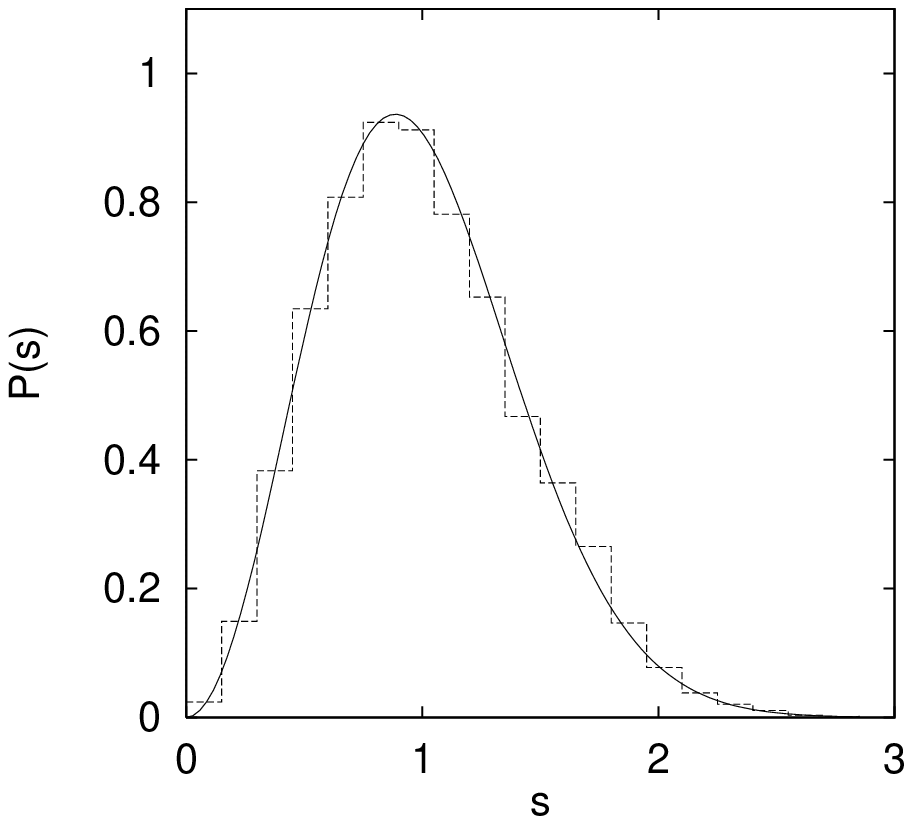}} &&
  \multicolumn{2}{c}{\epsfxsize=5cm\epsffile{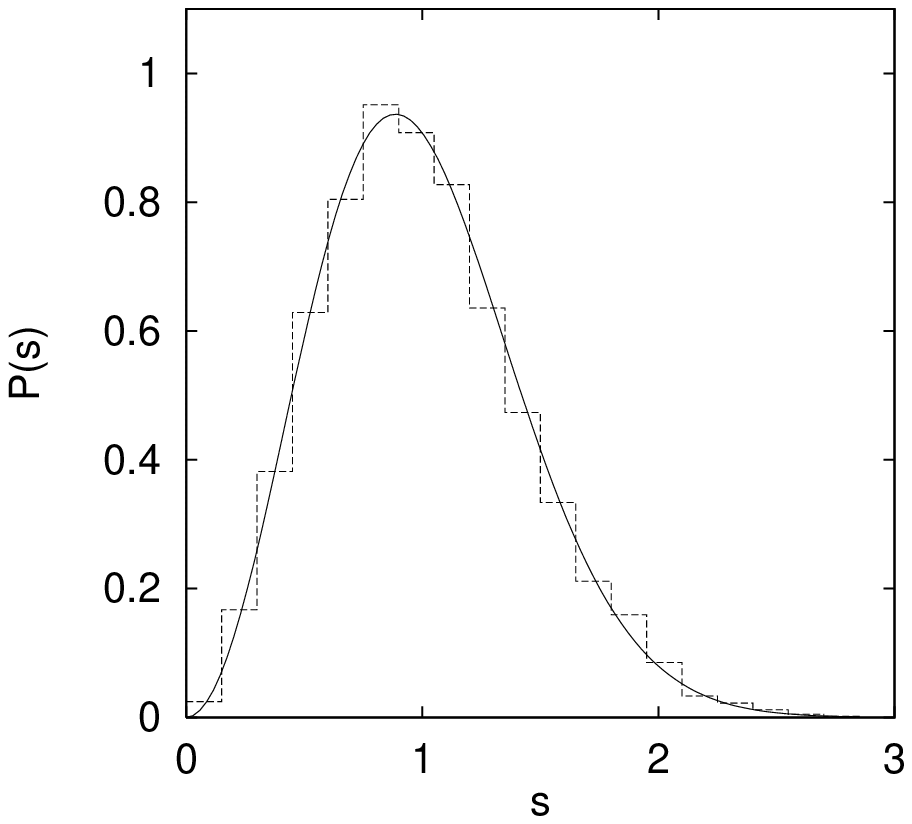}} \\
  \vspace*{-2mm}
  & $ma=0.05$ &&& $ma=0.05$ \\[2mm]
  \multicolumn{2}{c}{\epsfxsize=5cm\epsffile{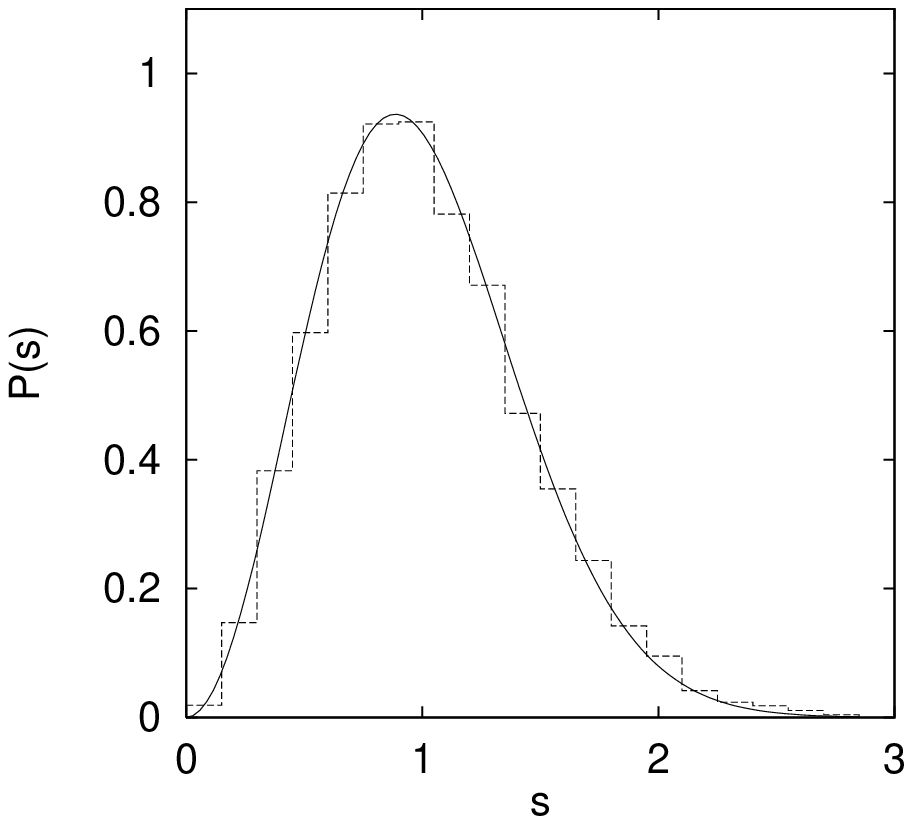}} &&
  \multicolumn{2}{c}{\epsfxsize=5cm\epsffile{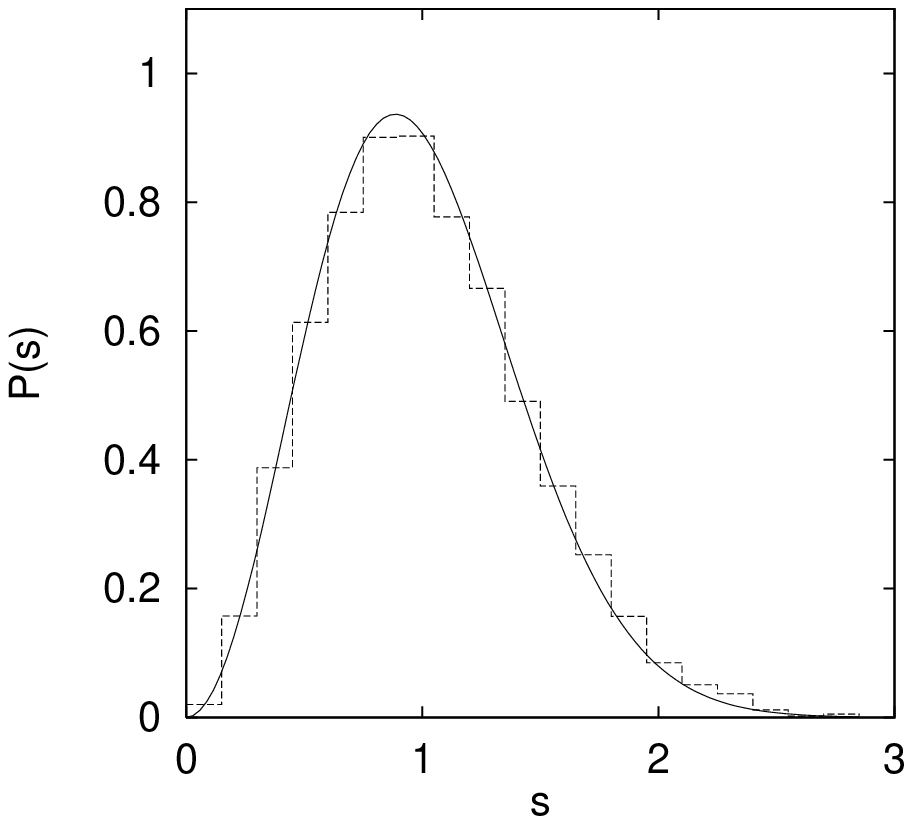}}
\end{tabular}
\end{center}
\vspace*{-1mm}
\caption{The nearest-neighbor spacing distribution $P(s)$ averaged
  over 10 independent configurations on a $6^3 \times 4$ lattice in
  full QCD (histograms) compared with the random-matrix result (solid
  lines).  Again, there are no changes in $P(s)$ across the
  deconfinement phase transition.}
\label{fig4}
\end{figure}

Figure~\ref{fig4} compares the nearest-neighbor spacing distribution
$P(s)$ of full QCD with the RMT result. No mass dependence is visible.
Again, in the confinement as well as in the deconfinement phase we
still observe agreement with the RMT result up to $\beta=8.0$ (not
shown).  The observation that $P(s)$ is not influenced by the presence
of dynamical quarks could have been expected from the results of
Ref.~\cite{Fox64}.  Those calculations, however, only apply to the
case of massless dynamical quarks.  Our results, as those of
Ref.~\cite{Hala95}, strongly indicate that massive dynamical quarks do
not affect $P(s)$ either.

No signs for a transition to Poisson regularity are found.  Thus, the
deconfinement phase transition does not seem to coincide with a
transition in the spacing distribution. For very large values of
$\beta$ far into the deconfinement region (not shown), the eigenvalues
start to approach the degenerate eigenvalues of the free theory, given
by $\lambda^2=\sum_{\mu=1}^4 \sin^2(2\pi n_\mu/L_\mu)/a^2$, where $a$
is the lattice constant, $L_{\mu}$ is the number of lattice sites in
the $\mu$-direction, and $n_\mu=0,\ldots,L_\mu-1$.  In this case, the
nearest-neighbor spacing distribution is neither Wigner nor Poisson.
However, it is possible to lift the degeneracies of the free
eigenvalues using an asymmetric lattice where $L_x$, $L_y$, etc. are
relative primes \cite{Verb97}. For large lattices, the
nearest-neighbor spacing distribution of the non-degenerate free
eigenvalues is then given by the Poisson distribution.  While it may
be interesting to search for a Wigner to Poisson transition on such
asymmetric lattices, it is unlikely that such a transition will
coincide with the deconfinement phase transition.
 
We do not believe that the absence of a signature for a transition
from Wigner to Poisson behavior at the deconfinement phase transition
is due to the finite lattice size.  Even for the small lattice size we
used, the agreement of $P(s)$ with the RMT curve is nearly perfect.
This leads us to believe that we should have seen some sign of a
transition if it existed in the thermodynamic limit.

\section{Conclusions}
\label{sec4}
 
We have searched for a transition in the nearest-neighbor spacing
distribution $P(s)$ from Wigner to Poisson behavior across the
deconfinement phase transition of pure gluonic and of full QCD.  Such
a transition exists, e.g., at the localization transition in
disordered mesoscopic systems \cite{mesosc}. In a Yang-Mills-Higgs
system a smooth transition along a Brody distribution was seen
\cite{Sala97}.  We observed no signature of a transition in our data
for the lattice Dirac matrix, neither for pure SU(3) nor for full QCD.
The data agree with the RMT result in both the confinement and the
deconfinement phase except for extremely large values of $\beta$ where
the eigenvalues are known analytically. Our analysis of full QCD with
two different quark masses showed no influence on the nearest-neighbor
spacing distribution.

Our findings are consistent with earlier studies of the chaotic
dynamics in classical lattice gauge theories which have shown that
lattice gauge theories are chaotic as classical Hamiltonian dynamical
systems \cite{Biro94}.  Also, it was found recently that the leading
Lyapunov exponents of SU(2) Yang-Mills field configurations indicate
that configurations in the deconfinement phase are still chaotic,
although less chaotic than in the strong coupling phase at finite
temperature \cite{Biro97}.

In hindsight, our results are not totally unexpected.  Temporal
monopole currents survive the deconfinement phase transition leading
to confinement of spatial Wilson loops.  Thus, even in the
deconfinement phase, the gauge fields retain a certain degree of
randomness.  It was shown that the relevant parameter that drives the
transition from Poisson regularity to chaos is of the order of the
mean level spacing $\Delta$ of the system \cite{Pand81,Guhr96}.  One
could search for a formal separation of the Dirac operator into a
regular and a chaotic contribution, $\FMSlash{D}=\FMSlash{D}_{\rm
  reg}+ (\alpha/\Delta)\, \FMSlash{D}_{\rm chaotic}$.  Then, a chaotic
perturbation of size $\alpha\,\raisebox{-1pt}
{$\stackrel{>}{\scriptstyle{\sim}}$}\,\Delta$ will generate a Wigner
distribution.  In future investigations, it would be interesting to
try to disentangle the corresponding contributions to the Dirac
matrix.

\section{Acknowledgments}

This work was supported in part by FWF project P10468-PHY and by DFG
grant We 655/11-2.  We thank T.S.\ Bir\'o, M.\ G\"ockeler, T.\ Guhr, 
E.-M.\ Ilgenfritz, M.I.\ Polikarpov, P.\ Rakow, A.\ Sch\"afer, and 
J.J.M.\ Ver\-baar\-schot for helpful discussions.

\end{document}